\documentclass[doublecol,figures]{epl2}
\title{Magnon pairing in quantum spin nematic}

\author{
M. E. Zhitomirsky\inst{1,2,3} \and H. Tsunetsugu\inst{2}
}
\shortauthor{M. E. Zhitomirsky and H. Tsunetsugu}

\institute{                    
\inst{1}Service de Physique Statistique, Magn\'etisme et Supraconductivit\'e,
UMR-E9001 CEA-INAC/UJF \\ 17 rue des Martyrs, F-38054 Grenoble cedex 9, France \\
\inst{2}Institute for Solid State Physics, University of Tokyo,
Kashiwanoha, 5-1-5, Chiba 277-8581 Japan \\
\inst{3}Max-Planck-Institut f\"ur Physik Komplexer Systeme, N\"othnitzer str.\
38,  D-01187 Dresden, Germany 
}

\pacs{75.10.Jm}{Quantized spin models, including quantum spin frustration }
\pacs{75.10.Kt}{Quantum spin liquids, valence bond phases and related phenomena}
\pacs{75.10.Pq}{Spin chain models}

\abstract{
Competing ferro- and antiferromagnetic exchange interactions may lead to the 
formation of bound magnon pairs in the high-field phase of a frustrated quantum 
magnet. With decreasing field, magnon pairs undergo a Bose-condensation prior 
to the onset of a conventional one-magnon instability. We develop an analytical 
approach to study the zero-temperature properties of the magnon-pair condensate, 
which is a bosonic analog of the BCS superconductors. Representation of the 
condensate wave-function in terms of the coherent bosonic states reveals the 
spin-nematic symmetry of the ground-state and allows one to calculate various
static properties. Sharp quasiparticle excitations are found in the nematic 
state with a small finite gap. We also predict the existence of a long-range 
ordered spin-nematic phase in the frustrated chain material $\rm  LiCuVO_4$ 
at high fields.
}

\date{\today}

\begin{document}

\maketitle

\section{Introduction}

Frustrated spin systems are interesting in general, as their zero-
and low-temperature properties are governed by quantum fluctuations.
These strong fluctuations may inhibit the formation of long-range
magnetic order, stabilizing instead a disordered spin liquid.
Following Anderson \cite{Anderson73}, much of the interest in the past
decades has been focused on the investigation of various quantum
spin-liquid states \cite{Fradkin, Misguich}.
Another interesting possibility is the appearance of unconventional
magnetic order characterized by partial breaking of the spin-rotational
symmetry $O(3)$. A specific example is provided by spin-nematic states,
which are analogous to the ordered phases of needle-like molecules
in liquid crystals. Spin-nematic phases have been discussed 
phenomenologically in \cite{Andreev84,Chandra91}, whereas 
identification of the relevant microscopic mechanism remains a challenging 
theoretical problem.

It was suggested long time ago \cite{Blume69} that a sizable biquadratic exchange
$({\bf S}_i\cdot {\bf S}_j)^2$ in magnetic insulators with $S\geq 1$
may stabilize a quadrupolar phase with vanishing sublattice magnetization
$\langle {\bf S} \rangle=0$, but a nonzero second-rank tensor, {\it e.g.},
$\langle  S_x^2\rangle\neq \langle  S_y^2\rangle=\langle  S_z^2\rangle$.
In most real compounds the biquadratic exchange is, however, rather small.
Recently, the interest in the biquadratic mechanism for the
spin-nematic order has been revived in connection with
the experiments on cold atomic gases \cite{Stenger98,Demler02}
and on the disordered magnetic material
$\rm NiGa_2S_4$ \cite{Nakatsuji05,Tsunetsugu06,Laeuchli06}.

In this Letter we explore an alternative mechanism
for the spin-nematic ordering based on competition between
ferro- and antiferromagnetic interactions in magnetic insulators
with an arbitrary value of the local spin including $S=1/2$.
Specifically, the mechanism operates in strong magnetic field
and is based on the formation of bound magnon pairs in
the fully polarized state \cite{Chubukov91,Kuzian07,Dmitriev09,Ueda09}.
This scenario has been studied numerically in a number of
works on the so-called ferromagnetic $J_1$--$J_2$ chain model
\cite{Heidrich06,Vekua07,Hikihara08,Sudan09,Heidrich09}
and its generalization to two dimensions (2D) \cite{Shannon06,Shindou09}.
Clear numerical evidence was found for the critical quadrupolar
correlations below the saturation field for the 1D case.
Note, however, that there is no true long-range 
order, nematic and otherwise, in one-dimensional quantum magnets 
at zero temperature. 
Therefore, a number of important questions on 
stability of the ordered nematic state  and its excitation spectra
are not answered by studying the purely 1D model.

The purpose of this Letter is to develop a simple   
analytical framework to treat the ground-state properties
and low-energy magnetic excitations in the phase with
a long-range spin-nematic order.
Our description of the condensate of bound magnon pairs
resembles in many aspects the BCS theory for the condensate
of bound electron pairs in superconductors.
In addition, we predict that a spin-nematic phase
must exist at high fields in the frustrated
chain material $\rm LiCuVO_4$ 
\cite{Gibson04,Enderle05,Banks07,Buttgen07,Schrettle08,Buttgen10}.

The phenomenon of magnon pair condensation has a close relationship 
to the old problem of particle versus pair-superfluidity in an attractive 
Bose gas \cite{Valatin58}. A common outcome is the density collapse 
prior to the pair-condensation transition \cite{Nozieres82}.
Unrestricted growth of the local magnon density in spin-1/2
antiferromagnets is cured by their hard-core repulsion.
In addition, reduced dimensionality of a spin subsystem
found in many real materials helps to stabilize
bound pairs and creates favorable conditions for their condensation.
Experiments on $\rm LiCuVO_4$ and other related
compounds in high magnetic fields
may, therefore, lead to the first observation of
such an exotic off-diagonal long-range order in solid-state systems.

\begin{figure}
\onefigure[width=7cm]{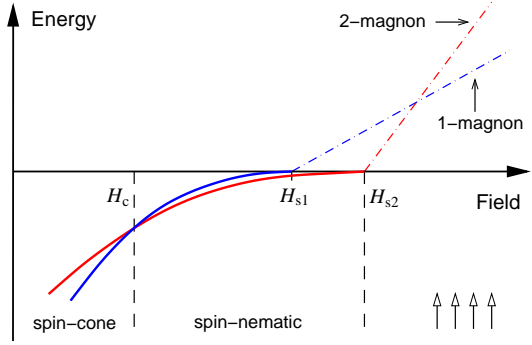}
\caption{(Color online) Energy-field diagram
for a frustrated quantum magnet close to the saturation field.
Dot-dashed lines show lowest one- and two-magnon states. Solid lines
represent the ground state energy for the one-magnon (spin-cone)
and the two-magnon (spin-nematic) condensate.
}
\label{energy}
\end{figure}

In order to demonstrate the occurrence of magnon pair condensation
at high magnetic fields we start with a general quantum Heisenberg
antiferromagnet on an $N$-site lattice
\begin{equation}
\hat{\cal H} = \frac{1}{2}\,\sum_{ i,{\bf r}}J({\bf r})\;
{\bf S}_i\cdot {\bf S}_{j} - H\sum_i S^z_i \ ,
\label{H0}
\end{equation}
where ${\bf r}={\bf r}_j-{\bf r}_i$.
In the following $S=1/2$ is set for definiteness.
In strong magnetic fields the Zeeman energy dominates over~the~exchange
interactions and stabilizes the fully polarized state $|0\rangle =
|\!\uparrow\uparrow\uparrow\!\ldots\,\rangle$.  This state is the vacuum
for single spin-flips or magnons
\begin{equation}
|1_{\bf q}\rangle = \frac{1}{\sqrt{N}} 
\sum_i e^{-i{\bf qr}_i}S_i^-|0\rangle
\end{equation}
with the excitation energy
\begin{equation}
\varepsilon_{\bf q} = H + \frac{1}{2}\, \sum_{\bf r} J({\bf r})
\bigl[e^{i\bf q r} - 1\bigr]  =
H + \frac{1}{2}\,(J_{\bf q} - J_0) \ ,
\label{E1k}
\end{equation}
where $J_{\bf q} = \sum_{\bf r} J({\bf r}) e^{i\bf q r}$.
In ordinary antiferromagnets spin-flips repel each other.
Then, once the band gap in $\varepsilon_{\bf q}$ vanishes at a certain
$\bf Q$, $J_{\bf Q}\equiv \min\{J_{\bf q}\}$,
an antiferromagnet undergoes a second-order transition
into a canted spin structure at the saturation field
\begin{equation}
H_{s1} = \frac{1}{2}\, (J_0-J_{\bf Q}) \ ,
\label{Hs1}
\end{equation}
as illustrated in Fig.~\ref{energy}.
The antiferromagnetic state below $H_{s1}$ can be regarded as a Bose-condensate
of single magnons \cite{Matsubara56,Batyev84}.

\section{Two-magnon bound states}

The conventional scenario for an antiferromagnetic transition 
in a strong magnetic field may change if some of the exchange bonds 
are ferromagnetic. In this case two spin-flips occupying the same 
bond with $J({\bf r})<0$ lower their interaction energy and may form 
a bound pair \cite{Chubukov91,Kuzian07}. To treat the bound state 
problem we follow the standard approach \cite{Wortis63,Hanus63,Mattis}
and define a general two-magnon state
\begin{equation}
|2\rangle = \frac{1}{2}\, \sum_{i,j} \, f_{ij} \, S_i^- S_j^- 
|0\rangle \ ,
\label{psi2}
\end{equation}
with $f_{ij} = f_{ji}$ being the magnon pair wave-function.
Separating the center of mass motion
$f_{ij} = e^{i{\bf k}({\bf r}_i+{\bf r}_j)/2} \,f_{\bf k}({\bf r})$
and calculating the matrix elements of $\hat{\cal H}$ for
states (\ref{psi2}) we obtain the Bethe-Salpeter equation
\begin{eqnarray}
& & \bigl( \varepsilon_2 - \varepsilon_{{\bf k}/2+{\bf q}}\!
- \varepsilon_{{\bf k}/2-{\bf q}}\bigr)\,
f_{\bf k}({\bf q}) = 
\label{E2} \\
& & \mbox{\quad} = \frac{1}{2N} \sum_{\bf p} \bigl( J_{\bf p+q}\! + J_{\bf p-q}\!
-  J_{{\bf k}/2+{\bf q}}\! -  J_{{\bf k}/2-{\bf q}}\bigr)f_{\bf k}({\bf p}),
\nonumber
\end{eqnarray}
where $f_{\bf k}({\bf q})$ is the Fourier transform of $f_{\bf k}({\bf r})$ 
and $\varepsilon_2$ measures the magnon pair energy relative to the energy 
of the ferromagnetically polarized state. The above equation extends 
the previous theories \cite{Chubukov91,Kuzian07,Ueda09} to an arbitrary
geometry of exchange interactions and with a trivial replacement of 
$\varepsilon_{\bf q}$ it also remains valid for an arbitrary value
of spin $S$.

\begin{figure}
\onefigure[width=7.5cm]{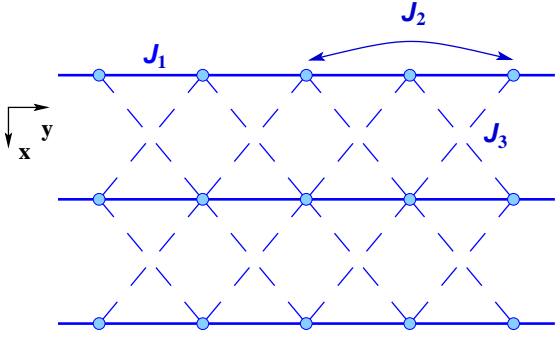}
\caption{(Color online) Two dimensional array of copper ions in $\rm LiCuVO_4$
with principal exchange couplings.
}
\label{bonds}
\end{figure}

While the subsequent theoretical consideration is entirely general,
we introduce now for illustration
a specific spin model shown in  Fig.~\ref{bonds}, which
is related to the quasi-1D helical antiferromagnet $\rm LiCuVO_4$.
This material consists of planar arrays of
spin-1/2 copper chains with a ferromagnetic nearest-neighbor
exchange $J_1<0$ and an antiferromagnetic second-neighbor
coupling $J_2>0$. Chains are linked by diagonal bonds,
whereas interplanar interactions are an order of magnitude smaller.
Neutron scattering measurements  provide
the following estimate for the exchange parameters in
$\rm LiCuVO_4$: $J_1=-1.6$~meV,
$J_2=3.8$~meV, and $J_3=-0.4$~meV \cite{Enderle05}.
The importance of quantum effects  in this material is revealed
by a small value of ordered  moments $\sim 0.3\mu_B$
in zero magnetic field \cite{Gibson04}.
The purely 1D $J_1$--$J_2$ model has been studied intensively in the past
\cite{Chubukov91,Heidrich06,Vekua07,Hikihara08,Sudan09,Heidrich09}
though no results exist for a realistic planar model.

To solve the integral equation (\ref{E2}) we expand $f_{\bf k}({\bf q})$
into lattice harmonics and obtain a finite algebraic system.
Bound states exist for $0.8\pi \leq k_y \leq \pi$ with the minimum
of $\varepsilon_2({\bf k})$ at ${\bf K} = (\pi,\pi)$. The binding energy
defined by $\varepsilon_2({\bf K}) = 2 \varepsilon_{\bf Q}-E_B$ is found
numerically to be  $E_B\approx 0.030J_2$. In the absence of bound states
the two-magnon continuum has a gap that is twice larger than 
the lowest one-magnon energy $\varepsilon_{\bf Q}=H-H_{s1}$.
The two gaps vanish, therefore, in the same magnetic field. 
When bound states are present, condensation of magnon pairs 
starts in a higher magnetic field: 
\begin{equation}
H_{s2} = H_{s1}  +  \frac{1}{2}\,E_B \ ,
\end{equation}
see Fig.~\ref{energy}. The condensation field for
$\rm LiCuVO_4$ is calculated to be $H_{s2} = 47.1$~T ($g=2$), whereas
the single magnon branch softens at $H_{s1} = 46.5$~T.
The relation $H_{s2}>H_{s1}$ holds up to $J_3\approx -0.6$~meV.
Hence, the conclusion about the magnon pair condensation in $\rm LiCuVO_4$
is rather robust and should not be affected by a possible
uncertainty in the experimental coupling constants \cite{Enderle05}.

We finish our analysis of the linear problem by presenting
the  wave-function of the lowest energy bound pairs
in the momentum representation:
\begin{equation}
f_{\bf K}({\bf q}) = \frac{\lambda J_2 \cos q_y}
{2J_2(1-\cos 2q_y) + 4J_3 \sin q_x\sin q_y + \delta \varepsilon }\,,
\label{fKq}
\end{equation}
where a numerical constant  $\delta\varepsilon = 0.130J_2$ is related
to the binding energy  and  $\lambda$ is the normalization factor.
In real space the function $e^{i{\bf Kr}/2}f_{\bf K}({\bf r})$ has 
the odd parity under the reflection $y\rightarrow -y$ and vanishes,
therefore, at ${\bf r}=0$.

\section{Coherent condensate of magnon pairs}
 
Below $H_{s2}$ the bound magnon pairs acquire
negative energy and start to condense. It is convenient at this point
to transform from spin-1/2 operators to the Holstein-Primakoff bosons:
\begin{equation}
S_i^z = \frac{1}{2} - a^\dagger_i a^{_{}}_i \ , \qquad
S_i^- = a_i^\dagger \sqrt{1 - a^\dagger_i a^{_{}}_i} \ ,
\label{HP}
\end{equation}
expanding subsequently square-roots to the first order in
$a^\dagger_i a^{_{}}_i$. 

The many-body state with a macroscopic number of the
lowest-energy pairs below $H_{s2}$ can be expressed as the coherent boson state
of the pair creation operator:
\begin{equation}
|\Delta\rangle = e^{-N|\Delta|^2/2}\,
\exp\Bigl[ \frac{1}{2}\,\Delta\sum_{ i,j }f_{ij}a^\dagger_i a^\dagger_j\Bigr]
|0\rangle \ .
\label{psi}
\end{equation}
Here $f_{ij} = e^{i{\bf K}({\bf r}_i+{\bf r}_j)/2}\,
f_{\bf K}({\bf r})$ is the wave-function of the lowest energy pairs
and $\Delta$  is the complex amplitude of the condensate.
The state (\ref{psi}) is a bosonic equivalent of the BCS pairing
wave-function for fermions and must be regarded as a variational 
ansatz, which can be further improved by taking into account pair-pair
correlations.

We use the ground-state wave-function (\ref{psi}) to compute
simple boson averages:
\begin{eqnarray}
&& \langle a_{\bf q}\rangle = 0\,, \quad
\langle a_{{\bf K}/2+{\bf q}} a_{{\bf K}/2-{\bf q}} \rangle
= \frac{\Delta f_{\bf K}({\bf q})}
{1 - |\Delta|^2 f^2_{\bf K}({\bf q})}\,,
\nonumber
\\[1mm]
&&
n_{{\bf K}/2+{\bf q}} =
\langle a^\dagger_{{\bf K}/2+{\bf q}} a^{_{}}_{{\bf K}/2+{\bf q}} \rangle =
\frac{|\Delta|^2 f^2_{\bf K}({\bf q}) }
{1 - |\Delta|^2 f^2_{\bf K}({\bf q})} 
\end{eqnarray}
as well as a more complicated four-boson correlator:
\begin{eqnarray}
&& \langle a^\dagger_{{\bf p}/2+{\bf q}} a^\dagger_{{\bf p}/2-{\bf q}}
a_{{\bf p}/2+{\bf q}'} a_{{\bf p}/2-{\bf q}'}  \rangle =
|\Delta|^2\,\delta_{{\bf p},{\bf K}} \nonumber
\\[1mm]
&&\mbox{} \ \ \ \ \ \times (1+ n_{{\bf K}/2+{\bf q}}+
n_{{\bf K}/2+{\bf q}'})\, \frac{ f_{\bf K}({\bf q})\, f_{\bf K}({\bf q}')}
{ 1- |\Delta|^4 f^2_{\bf K}({\bf q})\, f^2_{\bf K}({\bf q}')}
\nonumber
\\[1mm]
&& \mbox{} \ \ \ \ \ + |\Delta|^4\, (\delta_{{\bf q},{\bf q}'}+\delta_{{\bf q},-{\bf q}'}) \,
(1+ n_{{\bf p}/2+{\bf q}}+ n_{{\bf p}/2-{\bf q}})
\nonumber
\\[1mm]
&& \mbox{} \ \ \ \ \ \times \frac{
f^2_{\bf K}\bigl( \frac{{\bf p}-{\bf K}}{2} + {\bf q}\bigr)\,
f^2_{\bf K}\bigl( \frac{\bf p- K}{2} -{\bf q}\bigr) }
{1-|\Delta|^4\,
f^2_{\bf K}\bigl( \frac{{\bf p}-{\bf K}}{2} + {\bf q}\bigr)\,
f^2_{\bf K}\bigl( \frac{\bf p- K}{2} -{\bf q}\bigr) } \ .
\end{eqnarray}
From these one can derive various spin 
correlators. In particular, the absence of the single-magnon condensate 
$\langle a_{\bf q}\rangle = 0$
translates into 
\begin{equation}
\langle S_i^{x,y}\rangle = 0 \ ,
\end{equation}
which signifies a lack of a usual antiferromagnetic order parameter.
The transverse and longitudinal spin correlations are given in
the leading order by
\begin{equation}
\langle S_i^- S_j^+ \rangle \!\approx\! |\Delta|^2 \!\sum_l f^*_{il}f_{lj}, \quad
\langle \delta S_i^z \delta S_j^z \rangle \!\approx\! |\Delta|^2 |f_{ij}|^2
\label{spin_corr}
\end{equation}
with $\delta S_i^z = S_i^z -\langle S_i^z\rangle$. In accordance with the behavior
of the bound-state wave-function $f_{\bf K}({\bf r})$ the two correlators
decay exponentially as $|{\bf r}_i-{\bf r}_j|\rightarrow\infty$.
The transverse magnetic structure factor $S^\perp({\bf q})$
has no Bragg peaks and exhibits diffuse liquid-like spin
correlations with a characteristic shape in momentum space determined by
the bound-state wave-function:
\begin{equation}
\langle {\bf S}_{\bf q}^\perp\cdot {\bf S}_{-\bf q}^\perp \rangle \approx |\Delta|^2
\Bigl[ f^2_{\bf K}( \textstyle{\frac{\bf K}{2}} + {\bf q} )
+ f^2_{\bf K}( \textstyle{\frac{\bf K}{2}} \!-\! {\bf q})
\Bigr] \ .
\end{equation}
The longitudinal response $S^{zz}({\bf q})$  being
formally of the same order in $|\Delta|$ appears to be
much weaker than $S^\perp({\bf q})$ in the vicinity of the saturation field
$H_{s2}$.

The general definition of the spin nematic order parameter 
in the $O(2)$-symmetric case is 
\begin{equation}
Q^{\alpha\beta}_{ij} = \frac{1}{2}\, \langle S_i^\alpha S_j^\beta +
S_i^\beta S_j^\alpha \rangle - \frac{1}{2}\,\delta_{\alpha\beta}
\,\langle {\bf S}_i^\perp \cdot {\bf S}_j^\perp \rangle  \,
\end{equation}
where $i,j$ belong to a nearest-neighbor bond and $\alpha,\beta = x,y$.
The quadrupolar tensor $Q^{\alpha\beta}_{ij}$ acquires a nonzero expectation 
value in the presence of the  pair condensate:
\begin{equation}
Q^{xx}_{ij} + i Q^{xy}_{ij} = \frac{1}{2}\,
\langle S_i^+ S_j^+ \rangle \approx \frac{\Delta}{2}\, f_{ij} \ .
\end{equation}
The phase of the condensate amplitude $\Delta$ determines the orientation of
the spin-nematic director in  the $x$--$y$ plane. 
The director forms 
a periodic structure in the real space determined by the momentum $\bf K$.

Using explicit expressions for the spin correlators,
one can calculate the ground-state energy.
Below, we shall use for this purpose a complementary approach, which is
analogous to the Bogoliubov  method  in the theory of superconductivity
and yields the spectrum of quasiparticle excitations
simultaneously with the static properties.

\section{Mean-field approach}

The bosonic equivalent of the spin Hamiltonian
is obtained upon substitution of (\ref{HP}) into (\ref{H0}).
We restrict ourselves to quadratic and quartic terms in $a_i$'s
and define two types of mean-field  averages for each exchange bond:
\begin{equation}
\Delta_{ij} =\langle a_i a_j\rangle  \ , \qquad
n_{\bf r} = \langle a^\dagger_i a_j \rangle \ ,
\end{equation}
and the  magnon density
$n = \langle a^\dagger_i a_i \rangle$.
The anomalous correlator is further factorized as
$\Delta_{ij} = e^{i{\bf K}({\bf r}_i+{\bf r}_j)/2} \Delta_{\bf r}$.
Both $\Delta_{\bf r}$ and $n_{\bf r}$ are even real functions of $\bf r$
with a proper choice of gauge.
Performing the mean-field decoupling in the interaction term
we obtain a quadratic form, which is then diagonalized with the canonical
transformation.  This yields the energy of
one-magnon excitations
\begin{eqnarray}
&&
\varepsilon_{{\bf K}/2+{\bf q}} = \epsilon_{\bf q}
- \sum_{\bf r} J({\bf r}) ({\textstyle\frac{1}{2}}-n-n_{\bf r})
\sin{\textstyle\frac{1}{2}}{\bf Kr}\,\sin{\bf qr}, 
\nonumber \\
&&   \epsilon_{\bf q}=\sqrt{A_{\bf q}^2 -B_{\bf q}^2}\ , \ \ 
B_{\bf q} = \sum_{\bf r} J({\bf r}) \Delta_{\bf r} \cos{\bf qr} \ ,
\label{E1rk} \\
&& A_{\bf q} = H - \sum_{\bf r} J({\bf r}) ( {\textstyle \frac{1}{2}} - n -
n_{\bf r}) (1 - \cos{\textstyle \frac{1}{2}}{\bf Kr}\,\cos{\bf qr}).
\nonumber
\end{eqnarray}
In accordance with the exponential decay of spin correlations (\ref{spin_corr}),
the excitation spectrum acquires a gap in the presence 
of the magnon pair condensate.
The above expressions  are used to calculate bosonic averages
and to obtain a closed form of the self-consistent equations:
\begin{equation}
\Delta_{\bf r} =  \sum_{\bf q} \frac{B_{\bf q}}{2\epsilon_{\bf q}} \cos {\bf qr}\,, \ \ 
n_{\bf r} =  \sum_{\bf q} \frac{A_{\bf q}}{2\epsilon_{\bf q}} 
\cos ({\textstyle\frac{1}{2}}{\bf K}+{\bf q}){\bf r}\,.
\label{self_consist}
\end{equation}
In the limit $H\rightarrow H_{s2}$ one finds
$\Delta_{\bf r}\gg n_{\bf r}\sim \Delta_{\bf r}^2$, while
the linearized equation for $\Delta_{\bf r}$
transforms directly into the bound-state equation (\ref{E2}).

We have solved self-consistently the set of equations (\ref{E1rk}) 
and (\ref{self_consist}) and calculated the ground-state energy
for the spin model of Fig.~\ref{bonds}
assuming the same symmetry of bond variables $\Delta_{\bf r}$ and $n_{\bf r}$
as in the magnon pair coherent state (\ref{psi}).
With decreasing external field,
the pairs overlap more appreciably and at a certain point
give way to a conventional one-particle condensation.
Comparing the ground-state energy of the pair condensate with the energy of
a simple spin-cone structure we find the 
first-order transition at $H_c\approx 44.5$~T
as illustrated schematically  in Fig.~\ref{energy}. The spin-nematic state has the lowest
energy in a finite range of fields $H_c<H<H_{s2}$, which extends well
below the condensation field $H_{s1}$ for single magnons.

The ground-state energy calculation allows 
to determine the slope of the magnetization curve $M(H)$ at  $H_c<H<H_{s2}$.
In ordinary quasi-1D antiferromagnets, $M(H)$ deviates from a straight line
as $H\rightarrow H_s$ resembling the square-root singularity of a single 
quantum spin chain. 
Our mean-field calculations for the high-field nematic phase in
$\rm LiCuVO_4$ yield instead the slope $dM/dH \approx 0.38 M_{\rm sat}/J_2$,
which amounts to only 54\% of the slope of the classical magnetization curve.
The quantum corrections beyond the mean-field approximation 
should somewhat modify this value.
However, we expect them to be small
for the same reason as the suppression of critical fluctuations in
the BCS superconductors. Indeed, the size of the bound magnon pairs is rather large
extending to $\xi \sim 10$ interatomic spacing in the direction of chains.
Already for small magnon densities each bound pair is surrounded 
by many neighboring magnon pairs, which enforces the mean-field behavior.
Therefore, a distinct signature of the high-field nematic phase 
in $\rm LiCuVO_4$ will be a sharp
change in the slope of the magnetization curve.

\begin{figure}[t]
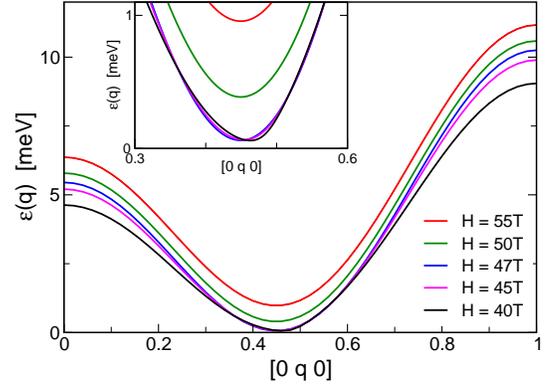

\onefigure[width=7cm]{dispersion.eps}
\caption{(Color online) Dispersion of a single-magnon branch
in $\rm LiCuVO_4$ 
in the fully polarized state ($H_{s2}\approx 47$~T) and 
in the state with the magnon pair condensate.
Field values for different curves from top to bottom are listed on the plot.
The inset shows the vicinity of the magnon gap.
}
\label{spectrum}
\end{figure}

The dispersion of one-magnon excitations found together with 
the ground-state energy in the self-consistent calculation 
is presented in Fig.~\ref{spectrum}. In the
fully polarized state at $H>H_{s2}\approx 47$~T, variation of 
the applied field results in an overall
shift of the magnon energy according to Eq.~(\ref{E1k}).
At $H=H_{s2}$ magnons have a small gap
$\Delta_g = H_{s2}-H_{s1} \approx 0.06$~meV.
The field dependence of $\varepsilon_{\bf k}$ changes drastically
in the presence of the magnon pair condensate. Decreasing the field modifies
the shape of $\varepsilon_{\bf k}$ but the excitation gap $\Delta_g$
remains practically unchanged, see the inset in Fig.~\ref{spectrum}.
The lowest field $H=40$~T used in Fig.~\ref{spectrum} is below the 
transition field $H_c$ into the spin-cone magnetic structure. 
The corresponding dispersion curve $\varepsilon_{\bf k}$ illustrates
that the spin-nematic state remains locally stable even below $H_c$.
This is in contrast with the previous scenario suggested for an 
attractive Bose gas, for which the pair condensate was assumed 
to become unstable due to a softening of the single-particle  
branch \cite{Valatin58,Nozieres82}.

The motion of the spin-nematic order parameter
provides an additional gapless branch of collective excitations, which
should yield a nonzero dynamical signal in the longitudinal channel. 
Transverse spin-spin correlations in the nematic phase
are dominated by unpaired magnons (\ref{E1rk}).

Finally, let us also comment on a low-field state at $H<H_c$. 
The considered scenario of a direct transition between the spin-nematic 
state and the conventional canted antiferromagnetic phase applies most 
certainly to the 2D model \cite{Shannon06}, which exhibits a magnetic 
ordering at ${\bf Q} = (\pi,0)$ in zero field. In the case of weakly 
coupled chains, the low-field phase might be more complicated than 
the simple conical spin-structure used above as an example.
The NMR measurements \cite{Buttgen07,Buttgen10} indicate 
that the intermediate-field phase, which is observed in $\rm LiCuVO_4$ 
above $8$~T \cite{Banks07}, has predominant longitudinal SDW-type 
correlations between local spins. This experimental finding agrees, 
in principle, with the numerical results for a single spin chain 
\cite{Vekua07,Hikihara08,Sudan09}.
Understanding the fate of such a 1D phase in the presence of 
interchain couplings
as well as its relation to the transverse nematic order remains 
an open theoretical problem.

To summarize, we have presented the analytical description for the 
Bose-condensate of bound magnon pairs in a frustrated quantum magnet
in high magnetic fields. The theory applies
to a number of real magnetic compounds with competing ferro- and 
antiferromagnetic interactions.
After the first version of the present work has appeared, we learned about
the experimental observation of a new phase in
$\rm LiCuVO_4$ in the field range 41--44~T \cite{Svistov10}.

\acknowledgments

We are grateful to M. Enderle, B. F\aa k, 
M. Hagiwara, and L. Svistov for stimulating discussions.
Part of this work has been performed within the Advanced Study
Group Program on ``Unconventional Magnetism in High Fields'' at
the Max-Planck Institute for the Physics of Complex Systems.
H.\,T.\ acknowledges support by Grants-in-Aid
for Scientific Research (No.~17071011 and No.~19052003) and
by the Next-Generation Supercomputing Project, Nanoscience Program,
MEXT of Japan.

\end{document}